\documentclass[twocolumn,superscriptaddress,longbibliography,aps,pra,preprintnumbers,10pt]{revtex4-2}
\usepackage[utf8]{inputenc}
\usepackage[urlcolor=blue,colorlinks=true,citecolor=blue,linkcolor=blue,pdfstartview={FitH},bookmarks=false]{hyperref}
\usepackage[T1]{fontenc}
\usepackage{graphicx}
\newcommand{\wymiar}{0.488\textwidth}
\begin{document}

\preprint{\textit{Submitted to:} \textbf{Physical Review B} (covering condensed matter and materials physics) published by American Physical Society (APS)}

\title{Electronic processes  occurring during  ultrafast demagnetization of cobalt\\ triggered by X-ray photons tuned  to Co L$_3$ resonance}

\author{Konrad J. Kapcia}
\email[e-mail: ]{konrad.kapcia@amu.edu.pl}
%\homepage[\mbox{ORCID ID}: ]{https://orcid.org/0000-0001-8842-1886}
\homepage[\mbox{ORCID ID}: ]{0000-0001-8842-1886}
\affiliation{\mbox{Institute of Spintronics and Quantum Information, Faculty of Physics}, Adam Mickiewicz University in Pozna\'n, 
Uniwersytetu Pozna\'{n}skiego 2, 61614 Pozna\'{n}, Poland}
\affiliation{\mbox{Center for Free-Electron Laser Science CFEL, Deutsches Elektronen-Synchrotron DESY}, Notkestr. 85, 22607 Hamburg, Germany}

\author{Victor Tkachenko}
\email[e-mail: ]{victor.tkachenko@xfel.eu}
\homepage[\mbox{ORCID ID}: ]{0000-0002-0245-145X}
\affiliation{\mbox{Institute of Nuclear Physics, Polish Academy of Sciences}, Radzikowskiego 152, 31-342  Krak\'ow, Poland}
\affiliation{European XFEL GmbH, Holzkoppel 4, 22869 Schenefeld, Germany}
\affiliation{\mbox{Center for Free-Electron Laser Science CFEL, Deutsches Elektronen-Synchrotron DESY}, Notkestr. 85, 22607 Hamburg, Germany}

\author{Flavio Capotondi}
\affiliation{Elettra-Sincrotrone Trieste S.C.p.A, 34149 Trieste, Basovizza, Italy}

\author{Alexander Lichtenstein}
\affiliation{European XFEL GmbH, Holzkoppel 4, 22869 Schenefeld, Germany}
\affiliation{University of Hamburg, Jungiusstr.  9, 20355 Hamburg, Germany}

\author{\mbox{Serguei Molodtsov}}
\affiliation{European XFEL GmbH, Holzkoppel 4, 22869 Schenefeld, Germany}

\author{Leonard Mueller}
\affiliation{Deutsches Elektronen-Synchrotron DESY, Notkestr.  85, 22607 Hamburg, Germany}

\author{Andre Philippi-Kobs}
\affiliation{Deutsches Elektronen-Synchrotron DESY, Notkestr.  85, 22607 Hamburg, Germany}

\author{Przemys\l{}aw Piekarz}
%\email[e-mail: ]{piekarz@wolf.ifj.edu.pl}
%\homepage[\mbox{ORCID ID}: ]{https://orcid.org/0000-0001-6339-2986}
%\homepage[\mbox{ORCID ID}: ]{0000-0001-6339-2986}
\affiliation{\mbox{Institute of Nuclear Physics, Polish Academy of Sciences},  Radzikowskiego 152, 31-342  Krak\'ow, Poland}

\author{Beata Ziaja}
\email[e-mail: ]{ziaja@mail.desy.de}
\homepage[\mbox{ORCID ID}: ]{0000-0003-0172-0731}
\affiliation{\mbox{Center for Free-Electron Laser Science CFEL, Deutsches Elektronen-Synchrotron DESY}, Notkestr. 85, 22607 Hamburg, Germany}
\affiliation{\mbox{Institute of Nuclear Physics, Polish Academy of Sciences},  Radzikowskiego 152, 31-342  Krak\'ow, Poland}

\date{\today}

\begin{abstract}
Magnetization dynamics triggered with ultrashort laser pulses has been attracting significant attention, with strong focus on the dynamics excited by VIS/NIR pulses. Only recently, strong magnetic response in solid materials induced by intense X-ray pulses from free-electron lasers (FELs) has been observed. The exact mechanisms that trigger the X-ray induced demagnetization are not yet fully understood. They are subject of on-going experimental and theoretical investigations. Here, we present a theoretical analysis of electronic processes occurring during demagnetization of Co multilayer system irradiated by X-ray pulses tuned to L$_3$-absorption edge of cobalt. We show that, similarly as in the case of X-ray induced demagnetization at M-edge of Co, electronic processes play a predominant role in the demagnetization until the pulse fluence does not exceed the structural damage threshold. The impact of electronic processes can reasonably well explain the available experimental data, without a need to introduce the mechanism of stimulated elastic forward scattering.
\end{abstract}

\maketitle

%%%%%%%%%%%%%%%%%%%%%%%%%%%%%%%%%%%%%%%%%%%%%%%%%
\section{Introduction}

Since its discovery \cite{Beaurepaire1996}, ultrafast demagnetization was studied mostly with lasers working in infrared regime \cite{Koopmans10,Kiril10,diffspin10,diffspin11,diffspin12,Sander17}. X-ray induced ultrafast demagnetization has become a topic of intense studies after the commissioning of the new generation light sources,  X-ray and XUV free electron lasers (FELs), see, e.g., \cite{Gutt10,Pfau12,Wang2012,Muller2013,Stohr2015,Wu2016,Willems2017,Chen2018,Schneider2020,Kobs2021,Kapcia2022}.  The FELs produce ultrashort, intense, coherent, and wavelength-tunable  X-ray pulses~\cite{Ackermann2007,Emma2010,Pile2011,Allaria2012}. Such pulses give an opportunity to study demagnetization by using X-ray magnetic circular dichroism (XMCD) effect, with FEL photons of energy tuned to a dichroic absorption edge of  a ferromagnetic element \cite{Coledge1,Coledge2,Hill96,Hannon88,Kiril10,Flavio13}. The principle of XMCD is, in particular, explored in the resonant magnetic small angle X-ray scattering (mSAXS) measurements ~\cite{Gutt10,Kobs2021,Riepp}, where magnetic response of irradiated systems can be probed on femtosecond timescales. The experiments which employed resonant magnetic scattering with photons tuned to M-absorption edge of cobalt, acting either as a pump or a probe, are  described in Refs. \cite{Gutt10,Pfau12,Muller2013,Schneider2020,Kobs2021}. The experiments which employed resonant magnetic X-ray scattering at L-absorption edge of cobalt are described in \cite{Wang2012,Wu2016,Chen2018}. Similar studies were also performed for nickel samples \cite{Stamm07,Hennes2021}.

The first theoretical explanation proposed for the observed loss of resonant magnetic scattering signal at M-edge of cobalt was proposed in \cite{Muller2013}. The mechanism considered was the perturbation of the electronic state within the magnetic sample during the first few femtoseconds of exposure leading to the atomic levels shifts and to the resulting  ultrafast quenching of the resonant magnetic scattering. However, the proposed mechanism was formulated in \cite{Muller2013} rather as a hypothesis and not proven explicitly there. In the following X-ray studies at Co L-edge \cite{Stohr2015,Wu2016,Chen2018}, the decreasing magnetic scattering signal was explained via stimulated elastic forward scattering in a simplified two-level atom model. However, this approach did not provide a full treatment of radiation damage caused by incoming X-ray photons. In particular, the two-level model approach could not account for the effect of electrons released during X-ray irradiation on electronic occupations within cobalt conduction band. Consequently, it did not accurately describe the magnetization dynamics triggered by photons tuned to M-edge of Co \cite{Kobs2021}. 

In Ref.~\cite{Kapcia2022}, we proposed a new modelling tool \textsc{XSPIN}, enabling a comprehensive nanoscopic description of electronic processes occurring in X-ray irradiated ferromagnetic materials. With this tool, we studied the response of Co/Pt multilayer system irradiated by an ultrafast XUV pulse  tuned to M-edge of Co (photon energy $\sim$ 60 eV), at the conditions corresponding to those of the experiment \cite{Kobs2021}. The \textsc{XSPIN} simulations showed that the magnetic scattering signal from cobalt decreased on femtosecond timescales due to electronic excitation, relaxation and transport processes, both in the cobalt and in the platinum layers. The signal decrease was stronger with the increasing fluence of incoming radiation, following the trend observed in the experimental data. Confirmation of the predominant role of electronic processes for X-ray induced demagnetization in the regime below the structural damage threshold, achieved with our theoretical study, was a step towards quantitative understanding of X-ray induced magnetic processes on femtosecond timescales. 

In this work, we apply the \textsc{XSPIN} model to describe the results of the experiment on resonant X-ray scattering with photons tuned to L$_3$-absorption edge of Co, performed at the Linac Coherent Light Source (LCLS) free-electron laser facility and  presented in Ref. \cite{Wu2016}. Although the electron kinetics following Co irradiation with X-ray photons of $\sim 778$ eV energy includes additional photoexcitation and relaxation channels such as inner-shell excitation and the resulting Auger processes, the electronic relaxation progresses in a similar way (through collisional processes) as after Co irradiation with 60 eV photons (M-edge case). Our purpose is to show that the collisional electronic relaxation is a universal mechanism that can explain ultrafast demagnetization of Co by X-ray pulses of low fluence, independently of X-ray photon energy.

In what follows, we will recall the details of the measurement performed in \cite{Wu2016} (Sec. \ref{section2:experiment}). Afterwards, we will discuss the application of the \textsc{XSPIN} model to model demagnetization induced by X-ray photons tuned to Co L$_{3}$-edge (Sec. \ref{section3:XSPIN}). 
We will then present the model results compared to experimental data (Sec. \ref{section4:results}). 
Finally, our conclusions will be listed (Sec. \ref{section5:conclusions}). 
In the Appendices, more details on our computational tools are presented.

%%%%%%%%%%%%%%%%%%%%%%%%%%%%%%%%%%%%%%%%%%%%%%%%%
% FIGURES
%%%%%%%%%
%%%%%%%%
% FIGURE 1
\begin{figure}[b!]
	\centering
    \includegraphics[width=\wymiar]{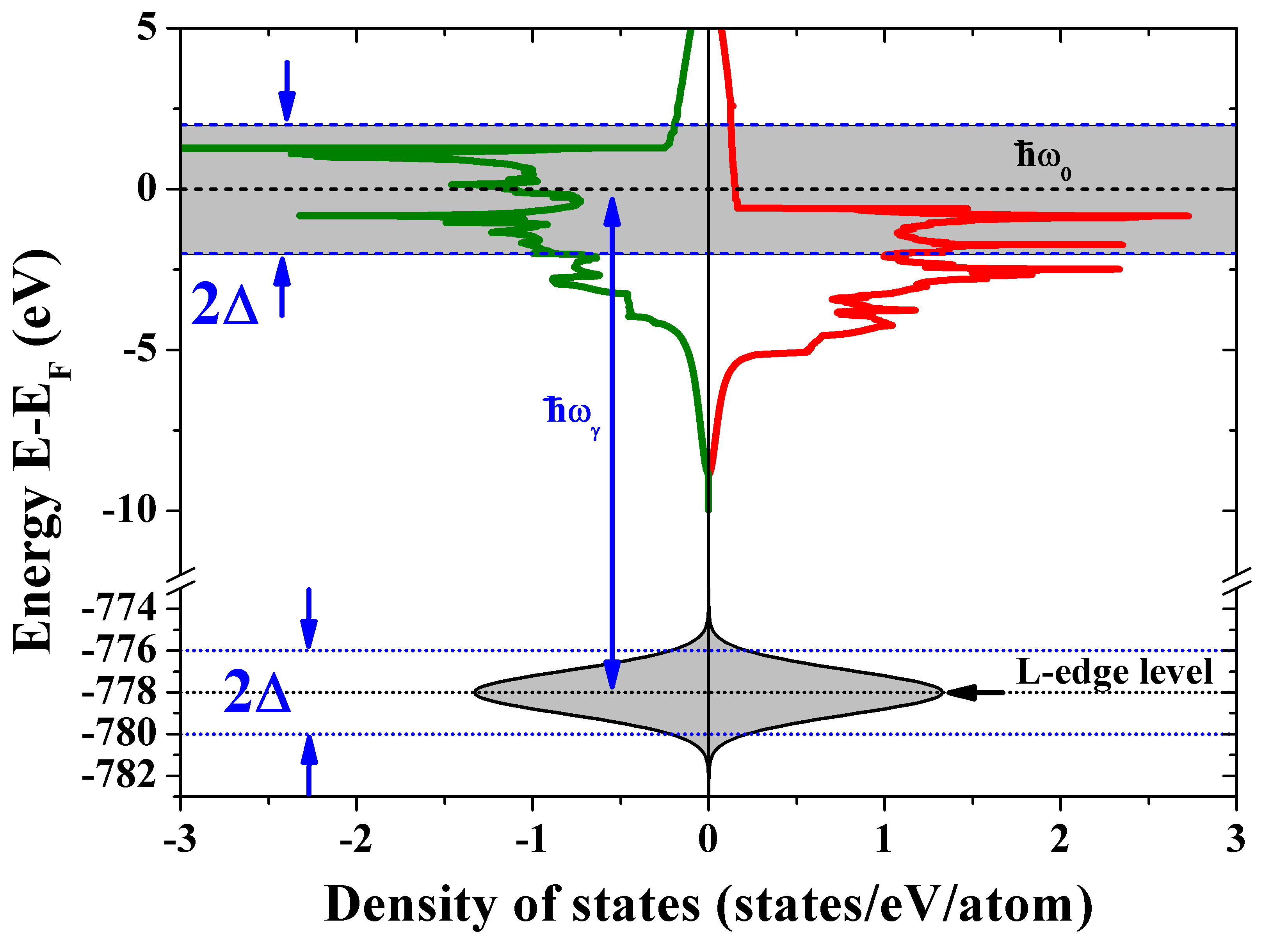}
    \caption{ \label{band}  
Calculated density of states (per atom) for equilibrium fcc cobalt, with schematic indication of the $2p$ band of cobalt, and of the probed region in its $3d$ band (the conduction band). The width of the $2p$ band is $2\Delta$. The density of states is shown for spin-up domain (red line) and for spin-down domain (green line).}
\end{figure}
%%%%%%%%

%%%%%%%%%%%%
%%%%%%%%%%%%%%%%%%%%%%%%%%%%%%%

\section{Resonant X-ray scattering experiment at LCLS facility}\label{section2:experiment}
%%%%%%%%%%%%%%%%%%%%%%%%%%%%%%%
In the experiment performed by Wu et al. at the LCLS facility \cite{Wu2016}, the Co/Pd magnetic multilayer system was used.  Its details are taken from Refs. \cite{Wang2012,Wu2016}. The multilayer system  consisted of Ta(1.5nm)Pd(3nm) [Co(0.5nm)Pd(0.7nm)]$_{40}$ Pd(2nm) layers fixed on the Si$_3$N$_4$ membrane. A similar magnetic system was also used in \cite{Chen2018}. For pumping and probing the system, the linearly polarized X-ray pulses of $778 \pm 0.1$ eV energy (monochromatized and tuned to the Co L$_3$-edge absorption resonance) were applied. Their duration was $50$ fs FWHM. The sample was covered by a radiation-opaque gold plate with a central hole of 1.45 $\mu$m diameter. XFEL pulses were focused onto the gold plate to a spot size of 10 $\mu$m FWHM. However, only a fraction of radiation  arrived on the sample, i.e., the fraction passing through the central hole in the plate. We have checked that the average pulse fluence in the aperture (i.e., the fraction of beam energy passing through the aperture, divided by the aperture size) was very similar to the average pulse fluence on the whole sample. The gold plate had also a few small holes outside the sample. The resulting X-ray diffraction patterns were recorded with a CCD detector. From the patterns, relative diffraction contrast of magnetic speckles was extracted for various values of the X-ray fluence. It was presented in Fig. 4(b) of Ref. \cite{Wu2016}. This quantity reflected the decrease of the magnetic scattering strength with increasing pulse fluence. Our model predictions will be later compared to it. 

%%%%%%%%%%%%%%%%%%%%%%%%%%%%%%%%%%%%%%%%%%%%%%%%%
\section{XSPIN model}\label{section3:XSPIN}

In order to follow X-ray induced magnetization change in irradiated magnetic material the \textsc{XSPIN} (details in Appendix \ref{sec:appA}) code has been developed \cite{Kapcia2022} as an extension of the hybrid code \textsc{XTANT}  \cite{Medvedev2013,Medvedev2018} (see Appendix \ref{sec:appB}). \textsc{XTANT} is an established simulation tool, enabling to study electronic and structural transitions triggered in solids by X-rays. The \textsc{XSPIN} code includes all the predominant processes occurring in solids as a result of X-ray irradiation, i.e., inner-shell and conduction band photoabsorption, Auger decay and collisional (impact) ionization, as well as electron thermalization.  In this work, we only consider electronic damage by X rays, assuming that the X-ray pulse fluence was too low to cause any structural damage resulting in atomic displacements. Although the dose absorbed per Co atom, needed to melt it thermally, seems rather low, $D_{melt}=0.54$ eV/atom, this melting criterium is not directly applicable to the femtosecond regime studied, as much longer time ($\sim$ few picoseconds)  is needed to fully melt Co after the absorption of   $D_{melt}$. Still, this dose gives a rough indication, at which value of the absorbed energy the processes leading to structural changes in Co can start to play a role. 

The foundations of the \textsc{XSPIN} code have been described in \cite{Kapcia2022}. There are two electronic subsystems -- with  spin-up and with spin-down electrons -- considered in the model.  Band structure for both subsystems is obtained from the density of states, $D_{\sigma}(\epsilon)$, calculated with the Vienna Ab initio Simulation Package (\textsc{VASP}) \cite{VASP,VASP1,VASP2,VASP3}. The energy levels in the low-energy electron fraction (here, containing electrons with energies less than 15 eV above the Fermi level) are determined from the total spin-polarized density of states $D_{\sigma}(\epsilon)$ calculated for fcc Co (for the experimental bulk value of the lattice constant, $a=3.545\ \text{\AA}$ \cite{Wang2019}), see Fig.~\ref{band}.   
The energy $E_{i,\sigma}$ of i-th level for spin-$\sigma$ electrons is then calculated from the equation, $i = \int_{-\infty}^{E_{i,\sigma}} d \epsilon D_{\sigma}(\epsilon)$~\footnote{% 
For example, for $512$ atom supercell, the numbers of levels between the bottom of the calculated (conduction) band and the cut-off energy of $15$ eV are $3832$ for spin-up electrons and $3792$ for spin-down electrons.}.
Note that here, $D_{\sigma}(\epsilon)$ is the total density of states for the system investigated (i.e., it is not normalized per the number of atoms in the system).
It is assumed that all electrons from the low-energy-electron fraction, both within the spin-up and the spin-down subsystems, stay in a common local thermal equilibrium. Therefore, their occupations on individual energy levels follow the Fermi-Dirac distribution, depending on the actual {\it common} electronic temperature and the {\it common} chemical potential. By the assumption of the mutual instant thermalization between both electronic subsystems, we implicitly include spin-flip processes into our model. 

After an X-ray pulse starts to interact with a solid material, electrons from spin-up and spin-down subsystems are released due to the photoabsorption processes. The excitation probabilities take into account the actual electronic occupations in the respective bands. If the photon energy is sufficiently high to trigger an electronic excitation from a core shell, a spin-up or spin-down electron can be excited from the shell. After the photoabsorption, the energetic photoelectron joins the non-thermalized high-energy electron fraction (i.e., here with the energies above 15 eV above the Fermi level). During the sequence of the following impact ionization events, the electron continuously loses its energy and, depending on its spin, ultimately joins either the spin-up or the spin-down subsystem of the  low-energy electron fraction. The high-energy electrons may collisionally excite secondary electrons, with the same or an opposite spin. The probability of such an excitation depends on the actual occupation of the spin-up and spin-down electron levels. Core holes relax via Auger decay. A band electron with the same spin fills the hole, while the Auger electron is chosen randomly, according to the actual distribution of spin-up and spin-down electrons.
In the code, at each time step an intrinsic averaging over $100000$ different Monte Carlo realizations of electron and holes trajectories is performed in order to calculate the average electronic distribution.

% FIGURE 2
\begin{figure}[t!]
	\includegraphics[width=\wymiar]{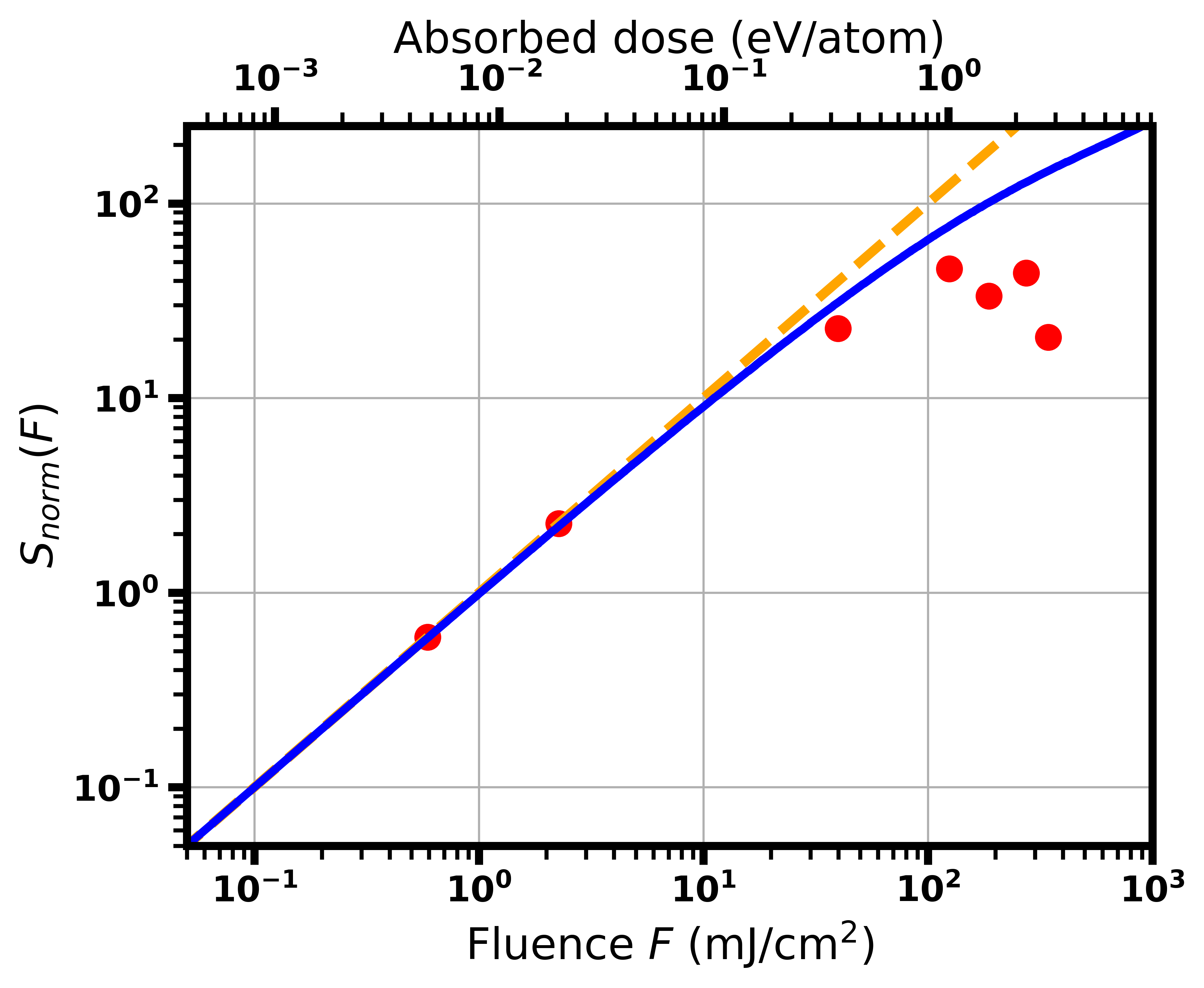}
\caption{
		\label{fig1} 
Normalized magnetic scattering efficiency, $S_{norm}(F)$ as a function of the average absorbed dose, also converted into the effective pulse fluence (incoming on the uppermost Ta layer).  
Predictions including the demagnetization (blue solid line), and predictions assuming no demagnetization, i.e., $M(t)=M(0)$ (orange dashed line) are shown for comparison. Experimental data (red circles) are taken from Ref. \cite{Wu2016}. The value of $\Delta=2$ eV was used \cite{Coledge1,Coledge2}.}
\end{figure}
%%%%%%%%
%%%%%%%%

The code \textsc{XSPIN} provides also the information on the strength of the resonant magnetic signal, scattered from the X-ray irradiated sample, i.e.,  the magnetic scattering efficiency $S(F)$. It is equal to the convolution of the incoming beam intensity and the actual magnetization of the sample \cite{Stoehr06,Schneider2020,Kapcia2022}: 
%%%%%%%%
\begin{equation}
 S(F) = \int_{-\infty}^{+\infty} dt \, I(t) M^2(t),
\end{equation}
%%%%%%%%
where the time-dependent magnetization $M(t)$ reflects the disparity between electronic populations at the resonant states in spin-up and spin-down electronic subsystems:
%%%%%%%%
\begin{equation}
M(t) = \sum_{\hbar \omega_{0} - \Delta}^{\hbar \omega_{0} + \Delta} \left[ N_{\uparrow}^{\textrm{hole}}(E_{i,\uparrow}) - N_{\downarrow}^{\textrm{hole}}(E_{i,\downarrow}) \right],
\end{equation}
%%%%%%%%
with $N_{\sigma}^{\textrm{hole}}(E_{i,\sigma})$ denoting the number of empty states at the $E_{i,\sigma}$ level. The \textsc{XSPIN} code calculates transient changes of $N_{\sigma}^{\textrm{hole}}(E_{i,\sigma})$  in response to a specific X-ray pulse for the probed energy levels within the Co $2p$ band, i.e., within the interval $\pm \Delta$  around the probed level $\hbar \omega_{0}$ (Fig.~\ref{band}). To explain it in more detail, the incoming X-ray photons of energy, $\hbar \omega_{\gamma}$, excite electrons from the $2p$ levels to the $3d$ band (the conduction band). The region in the $3d$ band, to which electrons are excited, then extends from $\hbar \omega_{0} - \Delta$ to $\hbar \omega_{0} + \Delta$, where $\hbar \omega_{0}$ is the difference between the photon energy and the position of L-edge: $\hbar \omega_{0} = \hbar \omega_{\gamma} - E_{\textrm{edge}}$, with $E_{\textrm{edge}}=778$ eV for L-edge of Co.  Here, $2\Delta$ is the $2p$  band width, which determines the number of states probed in the $3d$  band.
The number of holes within the probed interval of the conduction band then determines the strength of the recorded magnetic signal. Such a definition is a generalization of the standard definition of magnetization, where, for convenience, we calculate the difference between the unoccupied states (holes), instead of the difference between the occupied states (electrons).
%%%%%%%%%%%%%%%%%%%%%%%%%%%%%%%%%%%%%%%%%%%%%%%%%

% FIGURE 3
\begin{figure}[bh!]
%(a)  
\includegraphics[width=\wymiar]{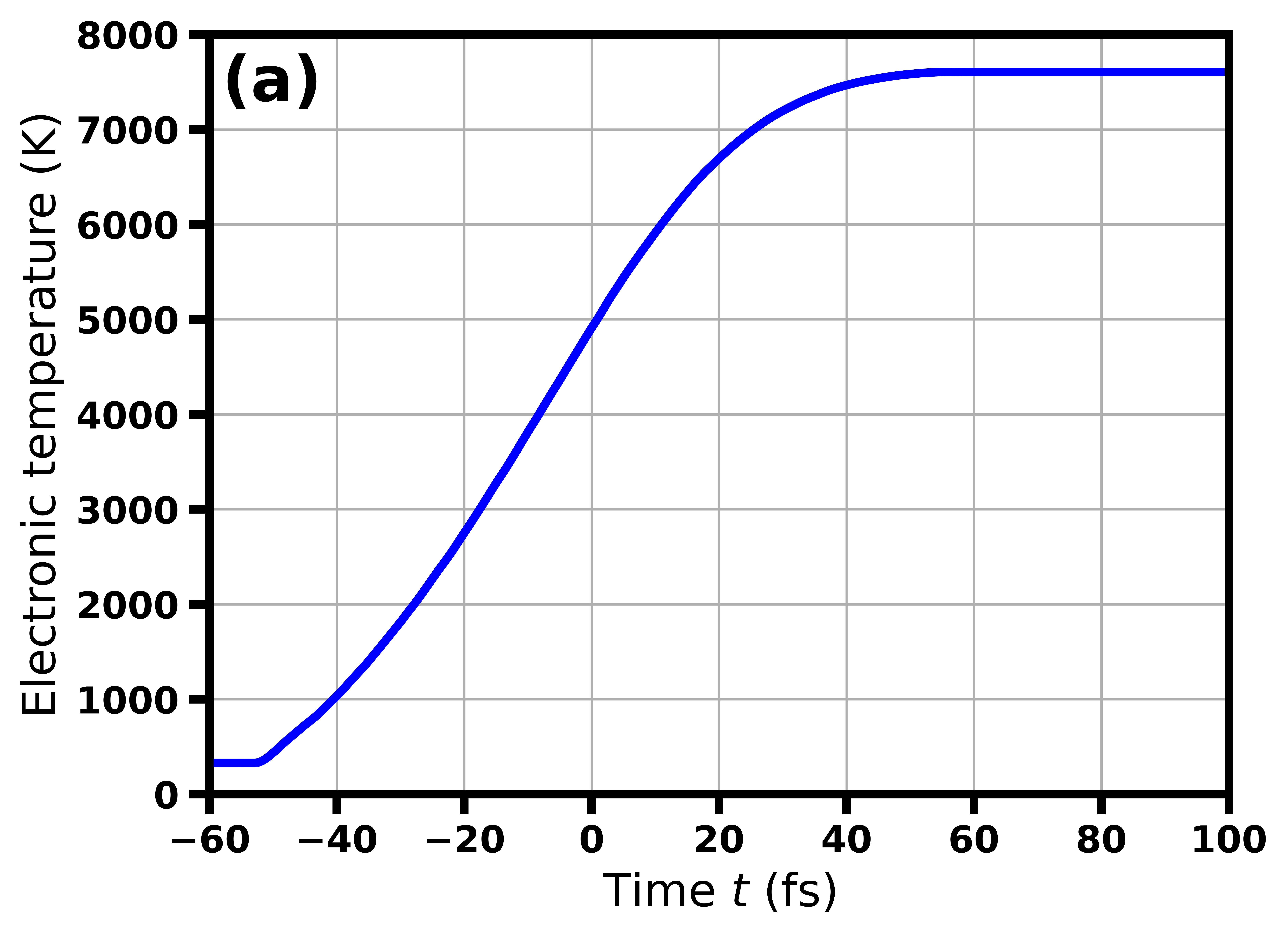}\\
%(b)  
\includegraphics[width=\wymiar]{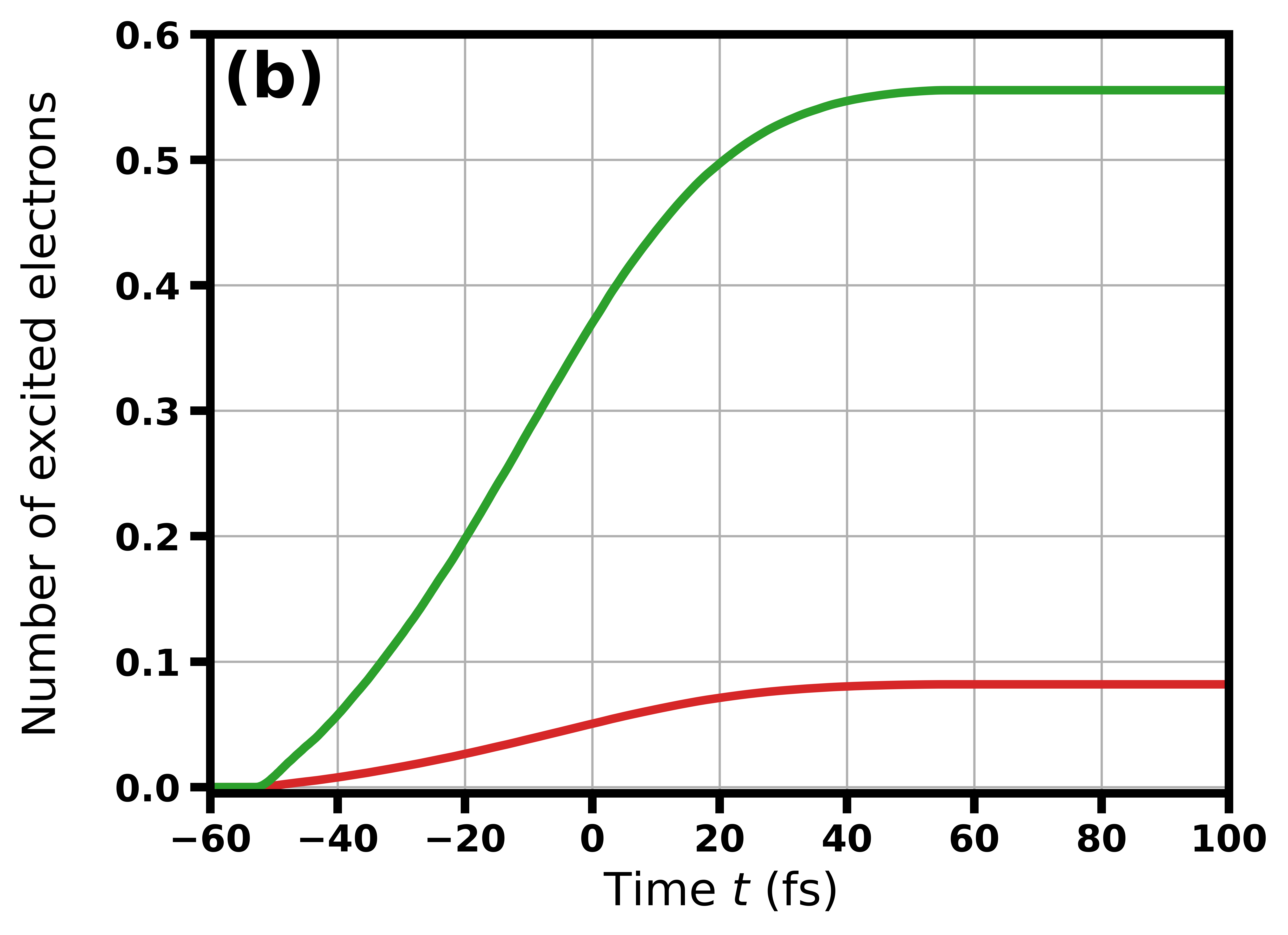}\\
%(c)  
\includegraphics[width=\wymiar]{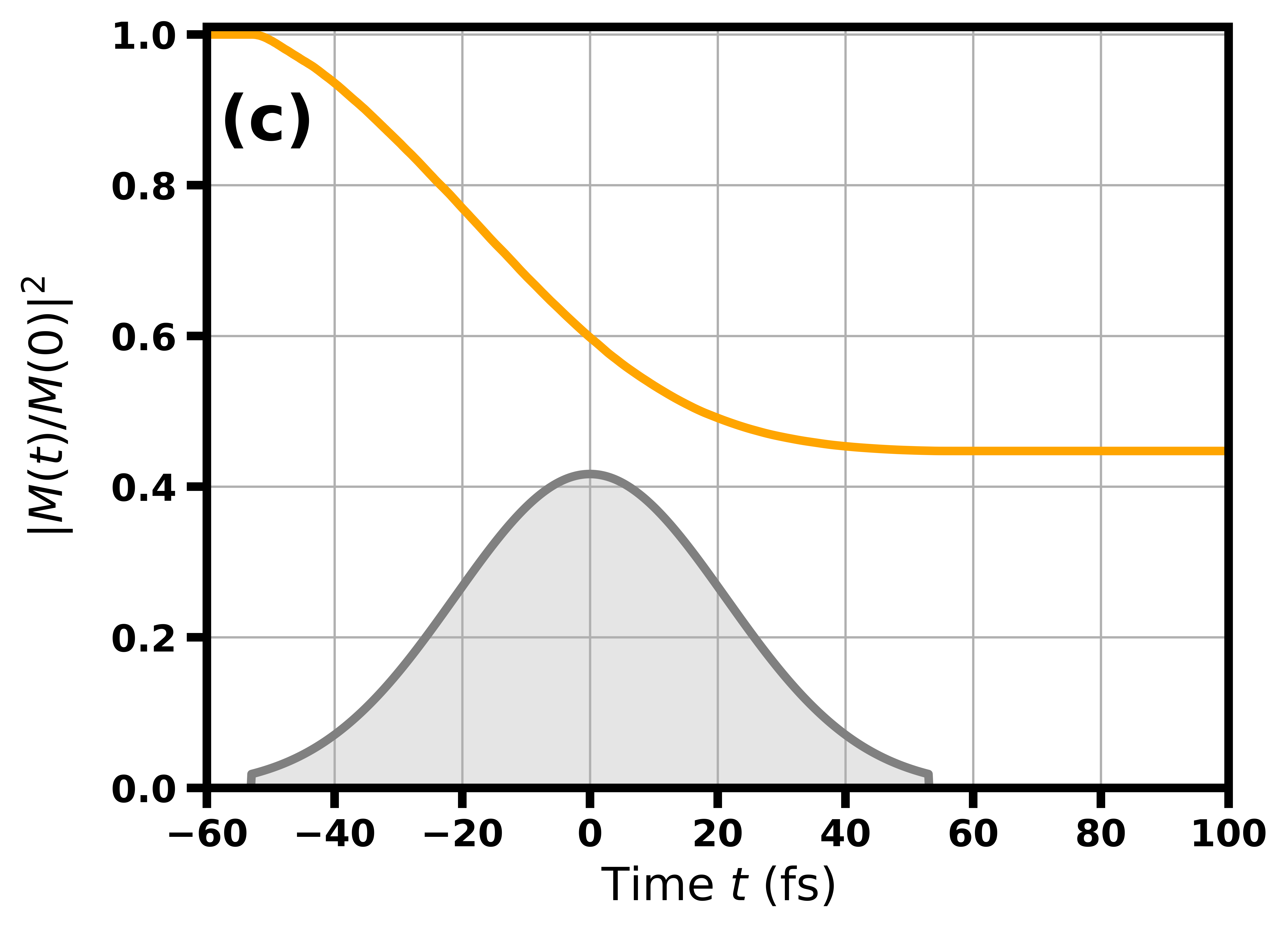}
\caption{	
		\label{fig2}
Temporal evolution of: (a) electronic temperature; (b) number of excited electrons (per atom) in spin-up and spin-down domains. Red line denotes the spin-up (majority spin) electron fraction and green line denotes the spin-down (minority spin) electron fraction; (c) magnetization, with schematically plotted X-ray pulse shape. The average absorbed dose used for this simulation was $0.93$ eV/atom. The value of $\Delta = 2.0$ eV was used for (c)  \cite{Coledge1,Coledge2}.}
\end{figure}
%%%%%%%%%%%%%%%%

\section{Results}\label{section4:results}

For the photon energy tuned slightly above L-edge of Co ($E_{\gamma}= 778.2$ eV),  the attenuation length of  X-rays in Co is $73.09$ nm, in Pd is $73.40$ nm, and in Ta is $99.89$ nm \cite{Henke1993}. If we compare these numbers with the overall size of the multilayer system, used in the experiment \cite{Wu2016}, we can conclude that the X-ray energy is absorbed to a large extent homogeneously in the sample. In addition, energetic electrons emitted as a result of  photoabsorption spread out on large distances within the sample (the electron ranges are $7.0$ nm for Co and $7.5$ nm for Pd). This additionally reinforces the homogeneity of the energy distribution in the sample. Therefore, we can estimate and use the same average effective absorbed dose for each of the Co layers. Assuming the homogeneous distribution of secondary electrons in the whole sample after the electron transport, the effective dose corresponds to the average dose absorbed per Co atom, needed to create the estimated average number of electrons in a Co layer. We estimated the linear conversion factor between the effective pulse fluence (incoming on the uppermost Ta layer) and the average  dose absorbed by Co atoms (which is an input parameter for \textsc{XSPIN}): $1000$ mJ/cm$^2$ corresponds to approximately $8.11$ eV/atom. This finding is in contrast to the \textsc{XSPIN} calculations in \cite{Kapcia2022}, where the energy absorption was strongly inhomogeneous in the multilayer system, even after the interlayer electron transport was included.

\textsc{XSPIN} simulations were performed for supercells with periodic  boundary conditions consisting of  $512$ atoms of Co.  This number of atoms ensured stability and convergence of the calculations. The atomic positions corresponded to the atomic positions in the equilibrium fcc cobalt.  Atoms were kept  frozen, i.e., no X-ray induced structural modifications in Co or Pd were taken into account. With the \textsc{XSPIN} predictions obtained for time-dependent magnetization $M(t)$ at various values of the absorbed dose, we calculated the magnetic scattering signal $S(F)$. Figure \ref{fig1} presents the results for the normalized magnetic scattering signal at $\Delta=2.0$ eV. This $\Delta$ value corresponds to the half of the FWHM of the Co L-edge resonance peak \cite{Coledge1,Coledge2,Hibberd2015,Guo2020}. Similarly as in \cite{Kapcia2022}, the normalized signal is defined as $S_{norm}(F)=S(F)\cdot F_0/S(F_0)$, where $F_0=0.01$ mJ/cm$^2$.  One can see initially a linear increase of $S_{norm}(F)$ with the pulse fluence. At the doses above 0.24 eV/atom (fluences higher than 30 mJ/cm$^2$), the curve starts to bend down and becomes non-linear, similarly as observed in \cite{Kapcia2022}. The results are compared with the experimental data from \cite{Wu2016} after converting the dose into pulse fluence arriving on Ta upper layer. There is a disagreement between  the data and theory predictions at higher fluence values. This is the regime above the structural damage threshold (0.54 eV/atom) where the  frozen-atom approximation may be not fully applicable, even at the short timescales considered.

In order to understand the processes behind the change of magnetic scattering signal observed, we analyzed the simulation results in detail. Figure \ref{fig2} presents temporal evolution of electron temperature, transient number of excited electrons (with energies above the Fermi level) per atom, and a typical shape of demagnetization curve $M^{2}(t)$ obtained for $\Delta=2.0$ eV, and normalized to its initial value. The temporal Gaussian profile of the X-ray pulse, with the duration of $50$ fs FWHM is also shown. The absorbed dose was in this case $0.93$ eV/atom. For this dose, one observes a $55$ \% decrease of $M^2(t)$, when compared with its initial value at $t=0$ fs. The decrease of the magnetization is related to the increase of the number of excited electrons (i.e., the electrons with the energy above the Fermi level; Fig.~\ref{fig2}(b) and the increase of electronic temperature [Fig.~\ref{fig2}(a)]. It is clearly seen [Fig.~\ref{fig2}(b)] that more spin-down (minority spin) electrons are excited than spin-up (majority spin) electrons. This is due to the 'asymmetry' between spin-up and spin-down bands (Fig.~\ref{band}), i.e., the different density of states for each spin orientation and the different energy level structure in each spin domain. For the spin-down domain, more energy levels are available above the Fermi level, and consequently, more electrons get excited there.  We have checked that although the number of spin-down and spin-up electrons is different, the energy absorbed in each of the electronic subsystems is comparable, as expected. 

% FIGURE 4
\begin{figure}[t!]
%(a) 
\includegraphics[width=\wymiar]{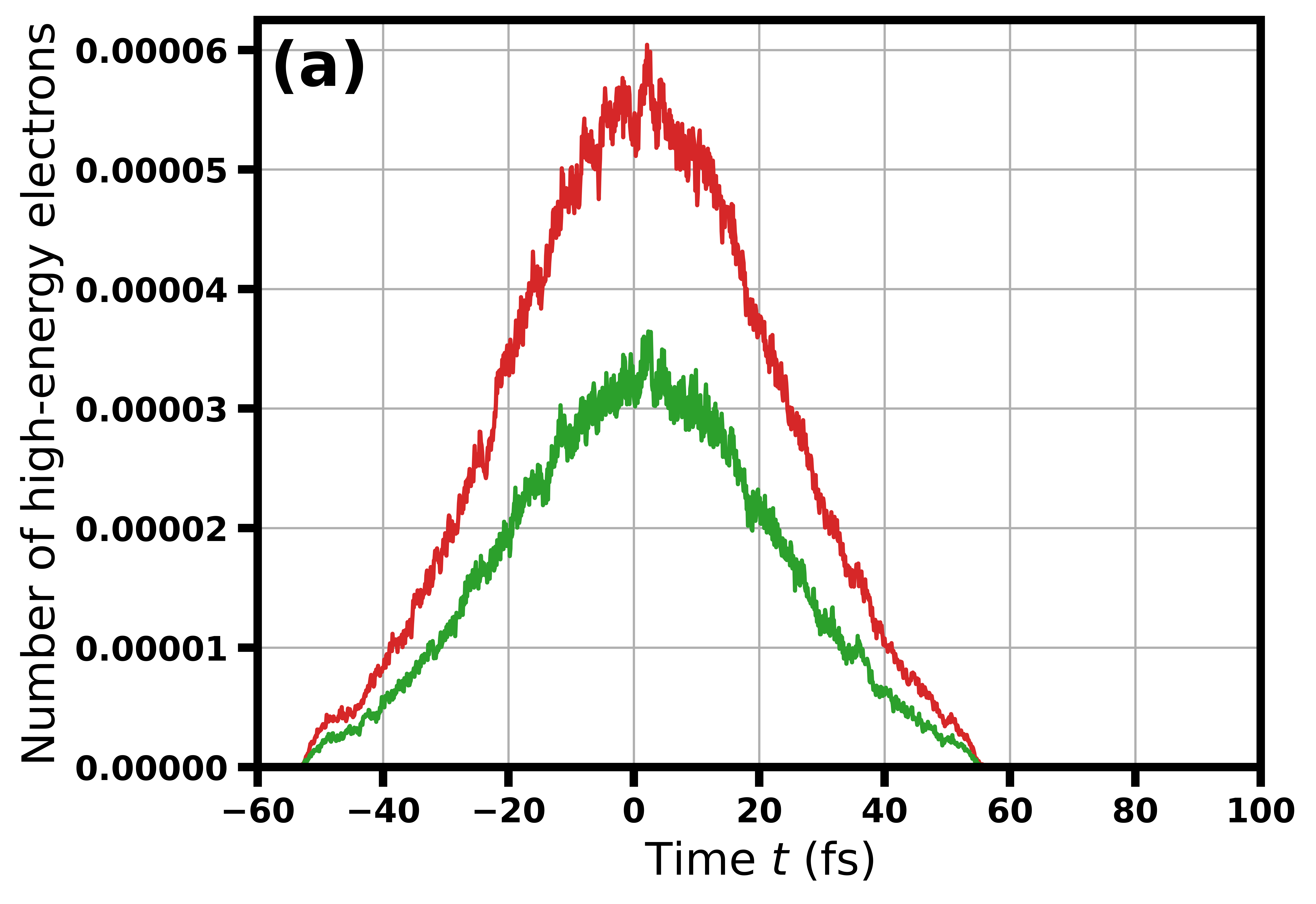}\\
%(b) 
\includegraphics[width=\wymiar]{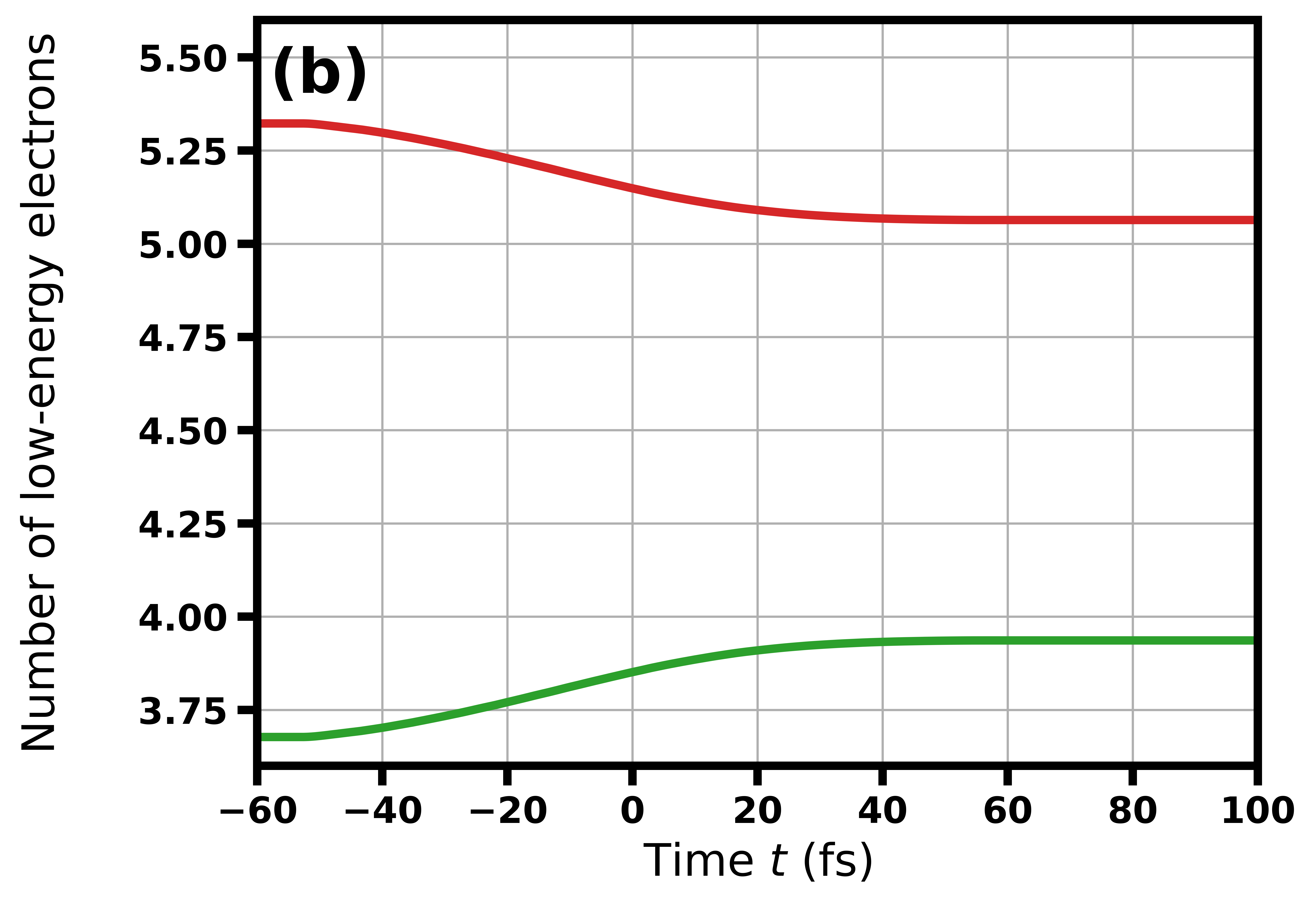} 
	\caption{	
		\label{fig3}
(a) Number of high energy electrons, and (b) number of low energy electrons (both per atom) as a function of time.  Red line denotes the spin-up (majority spin) electron fraction and green line denotes the spin-down (minority spin) electron fraction. The results were obtained for the average absorbed dose of $0.93$ eV/atom.}
\end{figure}

In Figure \ref{fig3}, we present time evolution of the total number of high-energy electrons (with energies above 15 eV) and low energy electrons (with energies below 15 eV) normalized per atom. The number of  high-energy electrons is relatively small and its evolution follows the intensity profile of the X-ray pulse. The transient number of high-energy spin-up electrons is larger than the number of high-energy spin-down electrons [Fig.~\ref{fig3}(a)]. Still, due to the above mentioned 'asymmetry' between spin-up and spin-down bands,  more electrons get excited to the low-energy spin-down domain during the collisional relaxation of high-energy electrons [Fig.~\ref{fig3}(b)]. As the result, the total number of spin-up electrons in the low-energy domain decreases and the total number of spin-down electrons respectively increases.  Ultimately, this drives the change of  Co magnetization,  depicted in Fig.~\ref{fig2}(c).
Note that, when electron cascading saturates, the value of $|M(t)|^2$ stabilizes, here within $\sim 55$ fs after the pulse maximum. As we can see in Fig.~\ref{fig3}(a), all high-energy electrons relax by this time into the band and they just occupy the levels above the Fermi level. Due to on-going exchange between the electronic system and lattice, the electrons will later loose more energy and finally thermalize with the lattice. However, this will happen on a picosecond time scale, i.e., outside the time window of the experiment and of the current simulation. Therefore, in our model, we do not treat the electron-lattice energy exchange \cite{Kapcia2022}).

%%%%%%%%%%%%%%%%%%%%%%%%%%%%%%%%%%%%%%%%%%%%%%%%%
\section{Conclusions}\label{section5:conclusions} 

In summary, we analyzed the role of electronic processes for ultrafast demagnetization in cobalt, triggered by X-ray photons tuned to L-edge of Co. The simulations performed with our computational tool \textsc{XSPIN} (which was already successful in describing magnetization dynamics triggered by photons tuned to M-edge  of Co \cite{Kapcia2022}), when compared to the L-edge data recorded a few years ago at the LCLS facility \cite{Wu2016}, proved a strong effect of electronic processes also for this case. More experimental data are needed for model validation in the 'destructive' fluence regime (i.e., for absorbed doses $> 0.54$ eV/atom). However, already now it is clear that the X-ray driven ultrafast rearrangement of electronic occupations within the magnetically sensitive bands of cobalt strongly impacts its magnetic properties. This observation opens a pathway towards quantitative control and manipulation of X-ray induced magnetic processes on femto- to picosecond timescales.

%%%%%%%%%%%%%%%%%%%%%%%%%%%%%%%%%%%%%%%%%%%%%%
\begin{acknowledgments}
B. Z. thanks A. Scherz and S. Parchenko for helpful discussions. K. J. K. thanks the Polish National Agency for Academic Exchange for funding in the frame of the Bekker programme (PPN/BEK/2020/1/00184). K. J. K. is also grateful for the funding from the scholarships of the Minister of Science and Higher Education (Poland) for outstanding young scientists (2019 edition, No. 821/STYP/14/2019). V. T. and B. Z. acknowledge the funding received from the Collaboration Grant of the European XFEL and the Institute of Nuclear Physics, Polish Academy of Sciences.
\end{acknowledgments}

\appendix

\section{Specific features of XSPIN model}\label{sec:appA}

\textsc{XSPIN}’s hybrid approach enables computationally inexpensive simulations of relatively large supercells (containing up to $1000$ atoms). The simulation scheme is based on the code \textsc{XTANT} (for details, see Appendix \ref{sec:appB}). The code treats all predominant electronic and hole core excitation and relaxation processes within an X-ray FEL irradiated sample, and follows sample's non-equilibrium and equilibrium evolution stage. Two electron distributions (with spins up and spins down) are evolved.

The  band structure levels in \textsc{XSPIN} are calculated with \textsc{VASP}  (Vienna Ab initio Simulation Package) code for materials in equilibrium \cite{VASP,VASP1,VASP2,VASP3}. The \textsc{VASP} is a code which enables high-precision density functional theory (DFT) calculations for various materials.  When applying this equilibrium calculation, we assume that the incoming X-ray pulses are not intense enough to cause any atomic displacements in the magnetic material during the exposure. We neglect also eventual shifts of electronic levels due to high electron temperature. As the nuclei positions are fixed, we can then use the ab initio density of states obtained for the material in equilibrium. 

A dedicated band structure module calculates the transient electronic occupations of the thermalized electrons. Electron occupation numbers, distributed on the transient energy levels, are assumed to follow the Fermi-Dirac distribution with a transient temperature and chemical potential evolving in time. The electron temperature and electron number changes due to the interaction of band electrons with X-rays and high energy electrons. We assume that all band electrons (both from the spin-up and from the spin-down fractions) undergo instantaneous thermalization at each time step. The intraband collisions, which lead to the electron thermalization, also include spin-flip collisional processes between spin-up and spin-down electrons. In such a way, the spin-flip processes are implicitly included in our model.

Non-equilibrium fraction of high-energy electrons and Auger decays of core holes are treated with a classical event-by-event Monte Carlo simulation. It stochastically models X-ray induced photoelectron emission from deep shells or from the valence band, the Auger decays, and the scattering of high-energy electrons. In the code, at each time step an intrinsic averaging over $100000$ different Monte Carlo realizations of electron (and core hole) trajectories is performed, in order to calculate the average electronic distribution which is then applied at the next time step.

More specific model details are listed below:
\begin{itemize}
\item[(i)]{In \textsc{XSPIN}, we assume that the X-ray fluences applied do not cause a significant structural damage to the material during or shortly after the XUV pulse, i.e., on $\sim 100$ fs timescales.  We give the justification below. First, a rigorous definition of structural damage threshold is difficult at the 100 fs timescale considered. The usual measure for a damage threshold in a metal is the threshold dose for its thermal melting. This dose for cobalt is estimated as $~0.54$ eV/atom. However, the thermal melting would require picosecond(s) to be completed. This time is needed for a transfer of a sufficient amount of energy from the electronic system to the lattice. At $100$ fs timescale, we can only use this threshold dose as an indicator when structural modifications can start to play a role. Second,  the usual timescale of atomic displacements during the structural transformation is longer than the femtosecond pulse duration, see, e.g., \cite{Tavella2017,Inoue2021,Tkachenko2021,Inoue2022}. 
These both observations guarantee a reasonable modeling accuracy even for the doses a few times higher than $0.54$ eV/atom, on $100$ fs timescales.  However, at higher absorbed X-ray doses or if the model should be applied at picosecond timescales (e.g., in order to follow the recovery of the magnetization), the possible atomic relocations should be taken into account. Such an extension of \textsc{XSPIN} is possible but it would require a significant modification of the anyway complex code, with much effort to be invested. Still, we plan this effort in future.}
\item[(ii)]{We assume that all band electrons (both from the spin-up and from the spin-down fractions) undergo instantaneous thermalization at each time step. The intraband collisions, which lead to the electron thermalization, also include spin-flip collisional processes between spin-up and spin-down electrons (cf. \cite{diffspin11}). In such a way, the spin-flip processes are implicitly included in our model.  Electron--ion coupling is neglected here, due to ultrashort timescales considered.
Note that the assumption of the instantaneous electron thermalization limits the applicability of the \textsc{XSPIN} to model X-ray irradiation with X-ray pulses of duration longer than the timescale of electronic thermalization. We have performed dedicated simulations with the \textsc{XCASCADE(3D)} code \cite{Lipp17} to investigate the timescale of electron cascading process in Co, which is comparable to the timescale of electron thermalization.  This indicates that the \textsc{XSPIN} model should not be applied for subfemtosecond X-ray pulses \cite{Kapcia2022}.}
\item[(iii)]{For the \textsc{XSPIN} analysis, we used average fluence values estimated by the experiment \cite{Wu2016}.  They were estimated, knowing the beam energy focused into a FWHM focal spot. We  assumed that the spatial profile of X-ray pulse in our simulations was flat-top, with an average fluence.  Assuming the homogeneous distribution of secondary electrons in the whole sample after the electron transport, the effective dose corresponds to the average dose absorbed per Co atom, needed to create the estimated average number of electrons in a Co layer. Note that attenuation lengths for Ta, Co, and Pd are following: $\lambda_{\textrm{Ta}} =99.89$ nm, $\lambda_{\textrm{Co}} =73.09$ nm, and $\lambda_{\textrm{Pd}} =73.40$ nm  \cite{Henke1993}, i.e., they are longer than the multilayer sample thickness ($54.50$ nm). Therefore, no in-depth volume integration of the signal was performed (cf. with \cite{Kapcia2022}).}
\item[(iv)]{\textsc{XSPIN} simulations were performed for the supercell containing $512$ Co atoms. As we consider fluences and timescales low enough not to cause atomic relocations, such number of atoms is sufficient to get a statistically reliable results. This expectation was confirmed by the preceding convergence tests of our results in respect to the size of the supercell (not shown).}
\item[(v)]{Interactions between magnetic domains in $(X,Y)$ plane are not included, consistently with the Stoner-Wolfarth model framework of a single magnetic domain \cite{Stoner47,Tannous08}, used here. Results from a simplistic model with periodic domains (not shown) indicate  that the details on domain structure in $(X,Y)$ plane should not significantly affect our results on $100$ fs timescales.}
\end{itemize}

\section{Modeling interactions of X-rays with solids using XTANT code}\label{sec:appB}

Modeling radiation damage in nanoscopic samples and solid materials has been performed for several years with various simulation techniques, e.g., \cite{Medvedev2013,Murphy2014XMDYN,Ho2020ARGONNE,Beyerlein2018CRETIN}. One of the tools is the hybrid code \textsc{XTANT} (X-ray-induced Thermal And Nonthermal Transitions) \cite{Medvedev2013,Medvedev2013PRB,Medvedev2017,Medvedev2018SciRep,Medvedev2018}. Using periodic boundary conditions, the  \textsc{XTANT} can simulate evolution of X-ray irradiated bulk materials. The code consists of a few modules dedicated to simulate various processes induced by the incoming X-ray FEL radiation: 
\begin{itemize}
\item[(a)]{The core of the \textsc{XTANT} model is a band structure module (in
\cite{Medvedev2013,Medvedev2017,Medvedev2018SciRep,Medvedev2018} based on transferable tight binding  Hamiltonian, in \cite{Lipp2022dftbp} replaced by the \textsc{DFTB+} code \cite{Hourahine2020dftbp}), which calculates the transient electronic band structure of thermalized  electrons and the atomic potential energy surface. The latter also evolves in time, depending on the positions of atoms in the simulation box, and is used to calculate the actual forces acting on nuclei.}
\item[(b)]{After the forces act on atoms, the atoms move. Their actual positions are propagated in time, using a classical molecular dynamics scheme. It solves Newton equations for nuclei, with the potential energy surface evaluated from the band structure module.} 
\item[(c)]{Electron occupation numbers, distributed on the transient energy levels, are assumed to follow  Fermi-Dirac distribution with a transient temperature and chemical potential  evolving in time. The electron temperature changes  due to the interaction of band electrons with X-rays and  high-energy electrons; or due to their non-adiabatic interaction with nuclei (through electron--ion scattering \cite{Medvedev2017}).} 
\item[(d)]{Non-equilibrium fraction of high-energy electrons and Auger decays of core holes are treated with a classical event-by-event Monte Carlo simulation. It stochastically models X-ray induced photoelectron emission from deep shells or from the valence band, the Auger decays, and the scattering of high-energy electrons.  In the code, at each time step an intrinsic averaging over $100000$ different Monte Carlo realizations of electron (and core hole) trajectories is performed, in order to calculate the average electronic distribution which is then applied at the next time step. Ballistic electrons are considered as high energy electrons. In the simulated bulk material, they propagate with the restriction of periodic boundaries.}
\item[(e)]{Electron--ion energy exchange can be calculated, using a non-adiabatic approach \cite{Medvedev2017}. This energy is transferred to atoms by the respective velocity scaling at each molecular dynamics step.}
\end{itemize}

\clearpage
%%%%%%%%%%%%%%%%%%%%%%%%%%%%%%%%%%%%%%%%%%%%%%

\bibliography{references-L-edge}

\end{document}